\documentclass{article}
\usepackage{amsmath,a4wide,latexsym,epsfig}
\newcommand{\intd}{\int \! d^4 x \;}
\newcommand{\intS}{\int \! d S \;}
\newcommand{\intSbar}{\int \! d \bar S \;}

\newcommand{\Ga} {\Gamma}
\newcommand{\Gacl} {{\Gamma_{\rm cl}}}

\renewcommand{\L}{{\cal L}}

\newcommand{\Tr}{{\rm Tr}}
\newcommand{\lambdabar}{{\overline\lambda}}

\newcommand{\epsilonbar}{{\overline\epsilon}}
\newcommand{\thetabar}{{\overline\theta}}
\newcommand{\etabar}{{\overline\eta}}
\newcommand{\chibar}{{\overline\chi}}

\newcommand{\fbar}{{\overline f}}

\newcommand{\alphadot}{{\dot\alpha}}
\newcommand{\betadot}{{\dot\beta}}

\def\df#1{\frac{\delta}{\delta#1}}

\def\pslash#1{{\setbox0=\hbox{$#1$}
  \rlap{\ifdim\wd0>.7em\kern.22\wd0\else\kern.1\wd0\fi /}#1}}

\def\brs{\mathbf s}
\newcommand{\mn}{{\mu\nu}}
\newcommand{\etabarbold}{{\mbox{\boldmath{$\etabar$}}}}
\newcommand{\etabold}{{\mbox{\boldmath{$\eta$}}}}

\begin{document}
%\begin{frontmatter}

\begin{flushright}
{\parbox{4cm}
{October 2002 \\
BN-TH-02-2002\\
hep-th/0211085 
}}
\end{flushright}

\vspace{0.5cm}

\begin{center}

{\Large \textbf{Non-renormalization theorems \\ \vspace{0.3cm} from extension to local 
coupling}\footnote{Talk presented at the \textit{10th International Conference
on Supersymmetry and Unification of
Fundamental Interactions} (SUSY 02), DESY, Hamburg, Germany, 17-23 June
    2002.}
} 

\vspace{0.5cm}

{\large {Elisabeth Kraus}}

\vspace*{0.2cm}
{\sl 
Physikalisches Institut, Universit\"at Bonn, Nu\ss allee 12,
D 53115 Bonn, Germany}

\end{center}
\vspace{0.5cm}
 \begin{abstract}
\noindent
The extension of  coupling constants  to
space time dependent fields, the local couplings, makes possible to
derive the non-renormalization theorems of supersymmetry by an algebraic
characterization of Lagrangian $N=1$ supermultiplets.
For super-Yang-Mills theories the construction implies
non-renormalization of the 
coupling beyond one-loop order. However,
 renormalization in presence of the
local gauge coupling is 
peculiar due to a new anomaly in one-loop order, which appears as an
anomaly of 
supersymmetry in the Wess-Zumino gauge. 
As an application we derive the closed all-order expression for the
gauge $\beta$ function and prove the non-renormalization of general N=2
supersymmetric  theories from a cancellation of the susy anomaly.

 \end{abstract}

\section{Introduction}

The extension of coupling constants to space-time dependent external
fields, i.e.~to local couplings, has been an important tool in
renormalized perturbation theory for a long time \cite{Bogo}. It has
been moved in the center of interest again  from a string point of
view since the couplings there are dynamical fields  and
 enter quite naturally as external fields the effective field theories
derived from 
string theories. It has been seen that local couplings
 in combination with  holomorphicity imply the non-renormalization
theorems
of supersymmetric  field theories
\cite{SHVA91,Louis91,SEI93,Louis94}.

Most of the results have been derived in the framework of
Wilsonian renormalization and by using the  Wilsonian effective
action. We have now started a rigorous and scheme-independent
construction of supersymmetric
field theories with local couplings.
In the Wess--Zumino model \cite{FLKR00}, in SQED \cite{KRST01} and
 in softly broken SQED \cite{KRST01soft} it was shown that local
couplings
 allow
the derivation of the
non-renormalization theorems \cite{FULA75,GSR79} as well as the
generalized non-renormalization theorem \cite{SHVA86}
from  renormalization properties of the
extended model in an algebraic context.  Most interesting is the
extension to super-Yang-Mills theories \cite{KR01,KR01anom}.
 There the classical analysis
applies in the same way  as for SQED, but in the procedure of renormalization 
one finds a new anomaly of supersymmetry, which implies finally
the non-holomorphic
contributions in the gauge $\beta$
function of pure super-Yang-Mills theories.

 \section{Local coupling in super-Yang-Mills theories} 

We consider super-Yang-Mills theories
 with a simple gauge group as for example $SU(N)$ in the Wess-Zumino gauge.
 In the Wess--Zumino gauge  the
vector multiplet consists of the gauge fields $A^\mu = A^\mu_a \tau_a$,
the gaugino fields $\lambda^\alpha= \lambda_a^\alpha \tau_a$ and their
complex conjugate fields $\lambdabar^\alphadot$ and 
the auxiliary fields $D= D_a \tau_a$. For constant coupling the 
 action
\begin{equation}
\Ga_{\rm SYM} = \Tr \intd\Bigl(
- \frac 1{4g^2}  G^{\mn}(gA) G_{\mn}(gA) +  i 
 \lambda^ \alpha
 \sigma^\mu_{\alpha\alphadot} D_\mu \lambdabar ^ {\alphadot} +
\frac 1 8  
D^2 \Bigr)\ ,
\end{equation}
with
\begin{eqnarray}
G^\mn (A)
& = & \partial^\mu A^\nu - \partial^\nu A^\mu +i  [A^\mu, A^\nu] \ ,\nonumber \\
D^\mu \lambda & = & \partial^\mu \lambda -  i g [A^\mu, \lambda]\ , 
\end{eqnarray}
is invariant under non-abelian gauge transformations and supersymmetry
transformations.
%\begin{equation} \begin{array}{rclcrcl}
%\delta_{ \alpha}A^\mu &= &  i \sigma^\mu_{\alpha \alphadot}
%\lambdabar ^{\alphadot }\ , & \qquad &\bar \delta_{ \alphadot}A^\mu &= & -
% i \lambda ^{\alpha }\sigma^\mu_{\alpha \alphadot}
%\ ,   \\
%\delta_{ \alpha} \lambda_{ \beta } & = &
%- \frac i2 \big( \epsilon _{\alpha \beta} D 
%- \sigma^{\mu \nu}_{\alpha \beta} G_{\mu \nu} \big)
%\ ,& \qquad & \bar \delta_{ \alphadot} \lambda_\alpha & =& 0   \ ,\\
%\delta_{ \alpha} \lambdabar_\alphadot &= & 0
%\ ,&\qquad& \bar \delta_{ \alphadot} \lambda_{ \betadot } & = &
%- \frac i2  \big(\epsilon _{\alphadot \betadot} D 
%- \bar \sigma^{\mu \nu}_{\alphadot \betadot} G_{\mu \nu} \big)
%\ ,  \\
%\delta_{ \alpha}D & = & 2 \sigma^\mu_ {\alpha
%\alphadot}
%D_\mu \lambdabar ^ { \alphadot} \ ,& \qquad & 
%\bar \delta_{ \alphadot}D & = & 2 
%D_\mu \lambda ^ { \alpha}\sigma^\mu_ {\alpha
%\alphadot} \ .
%\end{array}\end{equation}

In the Wess-Zumino gauge \cite{WZ74_SQED,DWit75} 
the algebra of supersymmetry transformation does not close on translations but 
involves an additional field dependent gauge transformation:
\begin{eqnarray}
\{  \delta_{\alpha }, \bar \delta_{\alphadot} \}
&= &
2 i \sigma^{\mu}_{\alpha \alphadot} (\partial^\mu +
\delta^{\rm gauge}_{A^\mu}) \ ,\nonumber \\
\{  \delta_{\alpha },  \delta_{\beta} \} &= &
\{ \bar \delta_{\alphadot }, \bar \delta_{\betadot} \} = 0\ .
\end{eqnarray}
Thus, on gauge invariant field monomials the supersymmetry algebra
takes their usual form,  and it is  possible to classify the Lagrangians
into the usual $N=1$ multiplets.

The Lagrangians of supersymmetric field theories are the $F$ or $D$
components of an $N=1$ multiplet. In particular one finds that the
Lagrangian of the super-Yang-Mills action is the highest component of
a chiral and an antichiral multiplet. Using a superfield notation in the chiral
representation  the chiral Lagrangian multiplet $\L_{\rm SYM}$ is given by
\begin{eqnarray}
\label{LYMmult}
\L_{\rm SYM} 
& = &  - \frac 12 g^2  \Tr \lambda^ \alpha \lambda_\alpha + 
\Lambda^ \alpha  \theta_ \alpha+ \theta^ 2 L_{\rm SYM} \ ,
\end{eqnarray}
with the chiral super-Yang-Mills Lagrangian $L_{\rm SYM}$ 
\begin{eqnarray}
%\Lambda_\alpha & = & - \frac i2 \Tr \bigl(g \sigma^{\mn
%\; \beta} _\alpha \lambda_ \beta G_{\mu \nu}(gA)  + g^2 D \lambda_ 
%\alpha \bigr ) \ ,\\
L_{\rm SYM} & = & \Tr \bigl(- \frac 14  G^{\mn}(gA) G_{\mn}(gA) +  i 
g \lambda^ \alpha
 \sigma^\mu_{\alpha\alphadot} D_\mu( g\lambdabar ^ {\alphadot}) + \frac 1 8  g^ 2
D^2 \nonumber \\
& & \quad
- \frac i 8 \epsilon^{\mu \nu \rho \sigma} G_{\mn}(gA) G_{\rho \sigma}
(gA) \bigr) 
\ . \label{LSYM}
\end{eqnarray}
By complex conjugation one obtains the 
   the respective antichiral multiplet $\bar {\cal L}_{\rm SYM}$.

The crucial point for the present considerations is the fact that the
topological term $\Tr\, G \tilde G$ appears in the supersymmetric
Lagrangians
 (\ref{LSYM})
and is
related to the kinetic term $- \frac 14 \Tr ( G G)$ via supersymmetry.
 Using a local, i.e.~space-time-dependent, gauge coupling
one is able to include the complete Lagrangian 
with the topological term into the action and the renormalization of
the gauge coupling will be related to the renormalization of the
topological term. 

For this purpose 
%We  extend the gauge coupling  $g$ to an external field
%$g(x)$ \cite{KRST01,KR01}.
% For maintaining at the same time supersymmetry the coupling has
%to be extended to a 
%supermultiplet. Therefore 
we introduce a chiral and an antichiral
field multiplet with dimension zero
$\etabold $ and $\etabarbold$:
\begin{equation}
\etabold= \eta + \theta \chi + \theta^2 f\ , \nonumber \\
\etabarbold= \etabar + \thetabar \chibar + \thetabar^2 \fbar \ ,
\end{equation}
and  couple them to the Lagrangian multiplets of the super-Yang-Mills action:
\begin{eqnarray}
\label{Gaclloc}
\Ga_{\rm SYM}
&= & -\frac 14
\intS \etabold \L_{\rm SYM} - \frac 14 \intSbar \etabarbold \bar \L_{\rm SYM}
\nonumber \\
& = & \intd \Bigl(\eta L_{\rm SYM} - \frac 12 \chi^\alpha \Lambda_\alpha - 
\frac 12 f g^2 \Tr \lambda^\alpha
\lambda_\alpha + \mbox{ c.c.}\Bigl)\ .
\end{eqnarray}
Now it is obvious that  the local coupling $ g(x)$ is related  to
  the real part  of the field $\eta $:
\begin{equation}
\eta + \etabar = \frac 1 {g^2(x)}\ .
\end{equation}
 The imaginary part of $\eta$ couples to the
topological term.
%$\Tr\, G^\mn \tilde G_\mn$.
  Thus, it takes the role of a space time dependent
$\Theta$ angle 
\begin{equation}
\eta - \etabar = 2 i \Theta\ .
\end{equation}

\newpage
In perturbation theory
 the dependence on the superfields $\etabold$ and $\etabarbold$ will
 be 
governed by the following properties of the classical action (\ref{Gaclloc}):
\begin{itemize}
\item
 $g(x)$ is the loop expansion parameter and the 1PI Green functions
satisfy in loop order  $l$ the  topological formula
\begin{equation}
N_g \Ga^ {(l)} = (N_A + N_{\lambda}+N_D + 2(l-1))\Ga^{(l)}\ .
\end{equation}
\item The $\Theta$ angle couples to a total derivative:
\begin{equation}
\Big(\df{\eta} - \df{\etabar}\Big) \Ga = [- \frac i4 G^{\mn} \tilde
G_{\mn}+ i \partial(\lambda \sigma \lambdabar)] \cdot \Ga\ ,
\end{equation}
and the classical action satisfies the identity
\begin{equation}
\label{holomorph}
\intd \Big(\df{\eta} - \df{\etabar}\Big) \Ga
= -i \intd \df{\Theta} \Ga = 0\ .
\end{equation}
\end{itemize}

\section{Invariant counterterms}

Invariant counterterms are the local field monomials which are
invariant under the classical symmetries and hold as counterterms to
the non-local loop diagrams in loop order  $l$. As such they are in
one-to-one correspondence with the possible UV divergences of the theory.

Constructing the invariant 
counterterms to $\Ga_{\rm SYM}$ with local coupling we find
from gauge invariance and
supersymmetry:
\begin{eqnarray}
\label{GaSYMct}
\Ga^{(l)}_{\rm ct, phys} 
&= & z^{(l)}_ {\rm YM}
\Bigl(-\frac 18
\intS \etabold^{1-l} \L_{\rm SYM} - \frac 18 \intSbar
\etabarbold^{1-l} \bar \L_{\rm SYM} \Bigr)
\nonumber \\
& = & z^{(l)}_{\rm YM}
\intd \Bigl( - \frac 14 (2g^2)^{l-1} G^\mn(gA) G_\mn(gA)  \nonumber
\\
& & \phantom{z^{(l)}_{\rm YM}
\intd \Bigl( -}
-\frac 18 (2g^2)^l (l-1) \Theta G^\mn(gA) \tilde G_\mn(gA) +
\ldots
\Bigr)\ .
\end{eqnarray}
For local coupling  
 the $\Theta$ angle couples to a total derivative only for $l=
 0$. Hence, using the identity (\ref{holomorph})
one finds that 
these counterterms as well as the respective UV divergences are
 excluded
in all loop orders except for one loop:
\begin{equation}
\label{ctholomorph}
z^{(l)}_{\rm YM} = 0 \qquad \mbox{for} \quad l \geq 2 \ ,
\end{equation}
i.e.~the $\Theta$ angle is not renormalized and 
 determines the
renormalization of the coupling via supersymmetry. The resulting
 counterterm action is the holomorphic effective action which has been
 stated already in \cite{SHVA86}:
\begin{eqnarray}
\label{Gaeff}
\Ga_{\rm eff} 
&= & 
-\frac 18
\intS (\etabold + z^{(1)}_{\rm YM}) \L_{\rm SYM} + \mbox{c.c.}
\end{eqnarray}

It is obvious from eq.~(\ref{GaSYMct}) that the  one-loop order is
special. Here the renormalization of the coupling is not related to
the renormalization of the $\Theta $ angle, and one has an arbitrary
counterterm $z^{(1)}_{\rm YM}$ and a corresponding UV divergence.
However, as 
  we will show in the next section, 
 the quantum  corrections to $\Tr\, G \tilde G$
induce  an anomaly of  supersymmetry in one-loop order. It  makes
quantization of super-Yang-Mills theories non-trivial such that
quantum results cannot be obtained from an effective action or by using
multiplicative renormalization. 
\newpage

\section{The anomalous breaking of supersymmetry}

To quantize supersymmetric field theories in the Wess-Zumino gauge
one includes the gauge transformations, supersymmetry
transformations and translations into the nilpotent BRS operator
\cite{White92a,MPW96a,HKS99}:
\begin{equation}
\brs \phi = \delta^{\rm gauge}_c + \epsilon^\alpha \delta_\alpha +
\bar \delta_\alphadot \epsilon^\alphadot - i  \omega^\nu \delta_\nu^T\ .
\end{equation}
The fields $c(x)$ are the usual Faddeev-Popov ghosts, $\epsilon^\alpha$,
$\epsilonbar^\alphadot$ and $\omega^\nu$  are
constant ghosts of supersymmetry and translations, respectively.

As for usual gauge theories BRS transformations are encoded in the
Slavnov-Taylor identity:
\begin{equation}
\label{ST}
{\cal S}(\Gacl) = 0
\end{equation}
with
\begin{equation}
\Gacl = \Ga_{\rm SYM} + \Ga_{g.f.} + \Ga_{\phi\pi} + \Ga_{ext.f.}\ .
\end{equation}
and the
Slavnov--Taylor identity 
expresses gauge invariance and supersymmetry of the classical action.

In addition to the Slavnov-Taylor identity the dependence on the
external fields $\etabold$ and $\etabarbold$ is restricted 
to all orders by the
identity (\ref{holomorph}) \cite{KR01}, which determines the
renormalization of the $\Theta$ angle.
%\begin{equation}
%\intd \Big(\df{\eta} - \df{\etabar}\Big) \Ga = 0\ .
%\end{equation}
%It is equivalent to the identity of the $\Theta$ angle
%(\ref{totalder}) of ordinary gauge theories.

In the course of renormalization 
the Slavnov-Taylor identity (\ref{ST}) has to be established
 for the 1PI Green functions to all
orders of perturbation theory.
From the quantum action principle one finds that the possible breaking
terms
are local in one-loop order:
\begin{equation}
{\cal S}(\Ga) = \Delta_{\rm brs} + {\cal O}(\hbar^2)\ .
\end{equation}
Algebraic consistency yields  the  constraint:
\begin{equation}
\brs_{\Gacl} \Delta_{\rm brs} = 0 \ .
\end{equation}
Gauge invariance can be established as usually, i.e.\ one has
\begin{equation}
{\cal S}(\Ga)\Big|_{\epsilon,\epsilonbar = 0}
 =  {\cal O}(\hbar^2) \ ,
\end{equation}
and  the remaining breaking terms depend on the supersymmetry ghosts
$\epsilon$ and $\epsilonbar$, and represent as such a breaking of
supersymmetry. 
Having gauge invariance established, the supersymmetry algebra closes
on translations. Using the supersymmetry algebra one obtains that the
remaining  breaking terms of supersymmetry are
variations of field monomials with the quantum numbers of the action
\cite{PSS80}:
\begin{equation}
\Delta_{\rm brs} = \brs_{\Gacl} \hat \Ga_{\rm ct}\ .
\end{equation}
However, not all of the field monomials in $\hat \Ga_{\rm ct} $
represent scheme-dependent counterterms of the usual form.
 There is one  field monomial in $\hat \Ga_{\rm ct} $, which
 depends on the logarithm of the gauge coupling, but whose BRS variation
 is free of logarithms:
\begin{eqnarray}
\label{Deltaanomalybrs}
\Delta^{\rm anomaly}_{\rm brs} & =   & \brs \intd \ln g(x) (L_{\rm SYM} + \bar
L_{\rm SYM}) \\
& = & (\epsilon^ \alpha \delta_\alpha + 
\epsilonbar^ \alphadot \bar \delta_\alphadot) 
\intd \ln g(x) (L_{\rm SYM} + \bar
L_{\rm SYM}) \nonumber\\
& = & \intd  \Bigl(i\; \ln g(x)  \bigl(\partial _\mu \Lambda ^
\alpha \sigma ^ \mu_{\alpha \alphadot} \epsilonbar^ \alphadot -
\epsilon^ \alpha \sigma^ \mu_{\alpha \alphadot}  \partial_\mu \bar \Lambda
^ {\alphadot} \bigr) \nonumber \\
& & \phantom{\intd} - \frac 12 g^ 2 (x) (\epsilon \chi + \chibar
\epsilonbar)
(L_{\rm SYM} + \bar L _{\rm SYM}) \Bigr)\nonumber\ .
\end{eqnarray}
%Here $L_{\rm SYM}$ and $\Lambda^\alpha$ are the $F$ and spinor
%component of the SYM multiplet $\L_{\rm SYM}$ (\ref{LYM}).
Indeed, due to the total derivative in the first line the breaking
$\Delta^{\rm anomaly}_{\rm brs}$ is free of logarithms 
for constant coupling and for any test with respect to the local
coupling. Moreover
  $\Delta^{\rm anomaly}_{\rm brs}$ satisfies
the topogical formula in one-loop order. 
Therefore 
$\Delta^{\rm anomaly}_{\rm brs}$  satisfies all algebraic constraints on the
breakings 
and can appear
as a breaking of the 
 Slavnov-Taylor identity in the first order of perturbation theory.

However, being the variation of a field monomial depending on the
logarithm of the coupling
 $\Delta^{\rm anomaly}_{\rm brs}$ cannot be induced by divergent
one-loop diagrams, which are all power series in the coupling.
Thus,
the corresponding
counterterm is not related to a naive contribution induced in the
procedure of subtraction and does not represent  a naive redefinition of
time-ordered 
Green functions. Indeed, it is straightforward to prove with algebraic methods
 that
the coefficient of the anomaly  is gauge and
scheme independent \cite{KR01}. 
 Therefore,
 $\Delta^{\rm anomaly}_{\rm brs}$ is an anomalous breaking of supersymmetry 
 in perturbation theory and we remain with
\begin{equation}
{\cal S}(\Ga) = r_\eta^{(1)} \Delta^{\rm anomaly}_{\rm brs} + {\cal
O}(\hbar^2)\ .
\end{equation}

 Evaluating the Slavnov--Taylor identity one can find an expression
for $r_\eta^{(1)}$ in terms of convergent loop integrals
\cite{KR01anom}.
 Using 
background gauge fields $\hat A ^\mu$ and Feynman gauge $\xi = 1$ 
 the anomaly coefficient is
explicitly related to insertions of the topological term and
 the axial current of gluinos into
  self energies of background fields:
\begin{equation}
\label{result}
 g^2 r^ {(1)}_\eta = - \frac 12 \Sigma^ {(1)}_{\eta-\etabar}(p_1,-p_1) \Big|_{ \xi=1} \ ,
\end{equation}
where $\Sigma_{\eta -\etabar}$ is defined by
\begin{eqnarray}
\label{Gatheta}
\Ga_{\eta - \etabar \hat A^\mu_a \hat A_b^\nu}(q, p_1,p_2)
&= &  \Bigl(\bigl[i\; \Tr \bigl(\partial( g^2 \lambda \sigma \lambdabar) - \frac 14 G^
{\mu\nu}\tilde G_{\mu \nu}(gA +\hat A) \bigr)\bigr] \cdot \Ga
\Bigr)_{\hat A_a^\mu \hat A_b^\nu}
(q,p_1,p_2)\nonumber \\ 
&   & { } = \; i \epsilon^ {\mu \nu \rho \sigma}p_{1 \rho } p_{2 \sigma} \delta_{ab}
\bigl(-2 + \Sigma_{\eta
- \etabar} (p_1,p_2)\bigr) \ .
\end{eqnarray}

From gauge invariance with background fields and local couplings 
$\Sigma_{\eta- \etabar}$ is unambiguously determined in perturbation
theory and not subject of renormalization.
Explicit evaluation  of the respective one-loop diagrams yields
\begin{equation}
\label{value}
r^ {(1)}_\eta = (-1 + 2)\frac {C(G)}   {8 \pi^2} =\frac {C(G)}   {8 \pi^2} \ ,
\end{equation}
where the first term comes from the axial anomaly of gauginos and
the second term from the insertion of the topological term \cite{KR01anom}.

We want to mention that the anomaly coefficient vanishes in SQED,
since one-loop diagrams to $\Ga_{\eta- \etabar  A^\mu  A^\nu}$ do not
exist. The anomaly coefficient also vanishes in $N= 2$ theories.
For $N=2$  super-Yang-Mills theories the analysis of the previous sections holds
in the same form, where $\L_{\rm SYM}$ and $\etabold$ are now $N=2$ chiral
multiplets. The algebraic analysis yields an anomaly of the same form
as  for $N=1$ theories, however, in the explicit evaluation the
 anomaly coefficient of $N=2 $ theories vanishes 
since  the two  fermionic fields just cancel the contribution arising
 from the topological term $\Tr\, G^\mn \tilde G_\mn$
(cf.~(\ref{Gatheta},
\ref{value})).

\section{Renormalization and the gauge $\beta$ function}

In the framework of perturbation theory
the anomaly of supersymmetry cannot be removed by a local
counterterm. However, one is able to proceed with algebraic
renormalization nevertheless by rewriting the anomalous breaking in
the form of a differential operator \cite{KR01}:
\begin{eqnarray}
\label{STanom1loop}
\Delta^{\rm anomaly}_{\rm brs}& = &
\intd \Bigl(g^ 6 r_\eta^{(1)}(\epsilon
\chi + \chibar \epsilonbar) \df {g^2} \nonumber  
\\ 
& & \qquad { }- 
i  r_\eta^{(1)} 
 \partial_\mu \ln g^ 2 \bigl( (\sigma ^ \mu \epsilonbar)^\alpha
\df {\chi^ \alpha}  + (\epsilon \sigma^ \mu )^ \alphadot \df {\chibar^
 \alphadot} \bigr)\Bigr) \Gacl
\end{eqnarray} 
Then one has
\begin{equation}
\big({\cal S} + r^{(1)}_\eta \delta {\cal S}\big) \Ga = {\cal
O}(\hbar^2)\ .
\end{equation}
For algebraic consistency one has to require the nilpotency properties
of the classical Slavnov-Taylor operator also for the extended
operator. Nilpotency  determines an algebraic consistent
continuation of (\ref{STanom1loop}). This continuation is not unique
but contains at the same time all redefinitions of the coupling
compatible with the formal power series expansion of perturbation
theory. 

As a result one finds an algebraic consistent
Slavnov-Taylor identity  in presence of the anomaly:
\begin{equation}
\label{STreta}
{\cal S}^ {r_\eta} (\Ga) = 0 \quad \mbox{and} \quad \intd \Bigl(\df{\eta} -
\df{\etabar} \Bigr)\Ga = 0\ ,
\end{equation}
where the anomalous part is of the form:
\begin{eqnarray}
{\cal S}^ {r_\eta} (\Ga) = {\cal S}(\Ga) & - & \intd \Bigl(g^ 4 \delta F(g^ 2)(\epsilon
\chi + \chibar \epsilonbar) \df {g^2} \nonumber \\ & &{ } \quad + 
i \frac {\delta F }{1+ \delta F}
 \partial_\mu g^ {-2}  \bigl( (\sigma ^ \mu \epsilonbar)^\alpha
\df {\chi^ \alpha}  + (\epsilon \sigma^ \mu )^ \alphadot \df {\chibar^
 \alphadot} \bigr)\Bigr)\Ga \ ,
\end{eqnarray}
with
\begin{equation}
\delta F(g^2)  = r^{(1)}_\eta g^2  + {\cal O}(g^4) \ .
\end{equation}
The lowest order term is uniquely fixed by the anomaly, whereas
the higher orders in $\delta F(g^2) $  correspond to the
scheme-dependent finite redefinitions of the coupling.

The simplest choice for $\delta F$ is given by
\begin{equation}
\label{Fmin}
 \delta F =  r_{\eta}^{(1)} g^2\ ,
\end{equation}
and another  choice is provided by
\begin{equation}
\label{FNSZV}
\frac{\delta F}{1 + \delta F} =  r_{\eta}^{(1)} g^2\ .
\end{equation}
As seen below, the latter choice gives the NSZV expression of the
gauge $\beta$ function \cite{NSVZ83,SHVA86}.

Algebraic renormalization with the anomalous Slavnov-Taylor operator (\ref{STreta}) is
performed in the conventional way.
In particular one can derive the $\beta$ functions from an algebraic
construction of the renormalization group equation in presence of
the local coupling. 
Starting from the classical expression of the RG equation
\begin{equation}
\kappa\partial _\kappa  \Gacl = 0
\end{equation}
we construct the higher orders of the RG equation
by constructing the  general basis of
symmetric differential operator with the quantum numbers of the action:
\begin{equation}
{\cal R} = \kappa \partial _\kappa + {\cal O}(\hbar) \ ,
\end{equation}
with
\begin{eqnarray}
& & \brs _\Ga^{r_\eta}{\cal R} \Ga -   {\cal R} {\cal S}^{r_\eta}(\Ga) =
 0
\ ,\label{RST} \\
& & \Big[\intd \Bigl(\df{\eta} - \df \etabar\Bigr), {\cal R} \Big] =
0\ .\label{Rholomorph}
\end{eqnarray}
The general basis for the symmetric differential operators
consists of the differential operator of the supercoupling
$\etabold$ and $\etabarbold$ and several field redefinition operators.
The differential operator of the coupling determines the $\beta$
function of the coupling, whereas the field redefinition operators
correspond to the anomalous dimensions of fields.

For the present paper
we focus on the operator of the $\beta$ function and neglect the
anomalous dimensions. For proceeding
we first construct  the RG operator ${\cal R}_{\rm cl}$, which is symmetric with respect to the classical
Slavnov-Taylor operator, i.e.\  we set $\delta F = 0$. In
a second step we extend it
 to a symmetric operator with respect to the full  anomalous
Slavnov-Taylor operator. 

Using  a superspace  notation we find:
\begin{equation}
\label{Rcl}
{\cal R}_{\rm cl} = -
\sum_{l \geq 1} \hat \beta_g^{(l)} 
\Bigl(\intS 
\etabold^{-l+1}\df {\etabold } + \intSbar \etabarbold^{-l+1}\df
{\etabarbold}
\Bigr) + \ldots\  ,
\end{equation}
and
\begin{eqnarray}
& & \brs _\Ga{\cal R}_{\rm cl} \Ga -   {\cal R}_{\rm cl} {\cal S}(\Ga) =
 0\ .
\end{eqnarray}
Evaluating then the consistency equation~(\ref{Rholomorph}) we obtain 
\begin{equation}
\hat\beta_g^{(l)} = 0  \quad \mbox{for } \quad l \geq 2 \ .
\end{equation}
These restrictions are the same restrictions as we have found for the
invariant counterterms of the super-Yang-Mills action (see 
(\ref{GaSYMct}) with
(\ref{ctholomorph})). Hence the only independent coefficient is the
one-loop $\beta$-function. 

 The one-loop operator ${\cal R}_{\rm cl}$ can be extended 
to a symmetric operator with
respect to the anomalous ST identity (\ref{RST}):
\begin{eqnarray}
\label{Rgen}
{\cal R} & =&  \hat \beta_g^{(1)} \intd g^3 (1 + \delta F(g^2)
)
\df{g} + \ldots \nonumber \\
& = &  \hat \beta_g^{(1)} \intd g^3 \big(1 + r_\eta^{(1)}g^2 + {\cal
O}(\hbar^2)\big) 
\df{g} + \ldots \ .
\end{eqnarray}
For constant coupling we find from (\ref{Rgen}) the closed expression
of the gauge $\beta$ function
\begin{equation}
\beta_g = \hat \beta_g^{(1)}  g^3 (1 + \delta F(g^2) ) =
\hat \beta_g^{(1)}  g^3 \big(1 + r_\eta^{(1)}g^2 + {\cal
O}(\hbar^2)\big) \ .
\end{equation}
Thus, the two-loop order is uniquely determined by the anomaly and the
one-loop coefficient, whereas higher orders depend on the specific
form one has chosen for the function $\delta F(g^2)$. In particular
one has for the minimal choice (\ref{Fmin}) a pure two-loop
$\beta$-function and for the NSZV-choice (\ref{FNSZV})  the NSZV
expression
\cite{NSVZ83} of the gauge $\beta$ function.

For $N=2$ super-Yang-Mills the anomaly vanishes and the classical
Slavnov--Taylor identity (\ref{ST}) can be extended to all orders. Then the
classical renormalization group operator (\ref{Rcl}) is a symmetric
operator to all orders and one gets from (\ref{Rholomorph})  a
pure one-loop contribution to the gauge $\beta$ function, i.e.,
\begin{equation}
{\cal S}(\Ga^{N=2}) =  0  \quad \Longrightarrow \quad \beta^{N=2}_g =
\hat \beta^{(1)}_g g^3 \ .
\end{equation}

\section{Conclusions}

The extension of the  coupling constants to an external fields is a 
  crucial step for deriving the non-renormalization theorems of
 supersymmetry in a scheme-independent way and independent from the
 usage of superspace methods. 
For super-Yang-Mills theories the extended model yields the
 non-renormalization of the coupling beyond one-loop order due to the
supersymmetry induced relation of the coupling with the $\Theta$
 angle. 

The non-renormalization of the $\Theta$-angle is the real new result gained
by the construction with local coupling. Using gauge invariance and the
property that the $\Theta$ angle couples to a total derivative in the
classical action higher order corrections a uniquely determined by
convergent expressions. In this way, local coupling in addition gives a
 simple proof of the non-renormalization of the Adler--Bardeen anomaly
\cite{AD69,ADBA69}
 also for non-supersymmetric theories.

For supersymmetric theories the non-renormalization of the $\Theta$
angle induces not only the non-renormalization theorem of the gauge
coupling but also  a supersymmetry anomaly in one-loop order. The
supersymmetry anomaly is the variation of a gauge-invariant field
monomial which depends on the logarithm of the local coupling. As such
it cannot be induced in the procedure of regularization, since loop
diagrams are power series in the coupling. Hence the anomaly found in
super-Yang-Mills with local coupling has the same properties as the
Adler-Bardeen anomaly: It is determined by convergent
one-loop diagrams and its coefficient is gauge- and scheme-independent. 

As an application we have constructed the renormalization group
equation and the gauge $\beta$ function. It was shown, that the
non-renormalization of the $\Theta$ angle first yields vanishing
coefficients for the $\beta$ function in $l\geq 2 $.  The
supersymmetry anomaly induces the 2-loop term in terms of the one-loop
coefficient and the anomaly coefficient. Higher order terms are
 scheme dependent and are determined by finite redefinitions of the
coupling. Hence, the non-holomorphic contributions in the $\beta$
function of pure super-Yang-Mills theories are generated by the
supersymmetry anomaly. Since $N=2$ super-Yang-Mills theories are not anomalous,
one can impose the classical Slavnov-Taylor identity and in this case
a pure one-loop gauge $\beta$ function is found.

The construction can be extended also to the matter part
\cite{KRST01,KR01}, 
and to
softly broken gauge theories
\cite{KRST01soft,KRST02soft}. Since the soft breakings are the lowest
components of Lagrangian multiplets, they are already included  in the
supersymmetric model with local coupling and thus 
softly broken supersymmetry appears as a natural extension of
supersymmetric theories with local coupling. 

\vspace{0.5cm}
{\bf Acknowledgments}

We thank R. Flume  
for many valuable discussions.

% Local Variables: 
% mode: latex
% TeX-master: "main"
% End: 

\end{document}